\documentclass[twocolumn,amsmath,amsfonts,showpacs,floatfix]{revtex4}
\usepackage{graphicx}
\usepackage{graphicx}
\usepackage{epstopdf}

\begin{document}
\title{Entanglement in a three spin system controlled by electric and magnetic fields}

\author{Jakub {\L}uczak and Bogdan R. Bu{\l}ka}
\affiliation{Institute of Molecular Physics, Polish Academy of Sciences, ul. M. Smoluchowskiego 17, 60-179 Pozna{\'n}, Poland}

\date{\today}

\begin{abstract}
We study the effect of electric field and magnetic flux on spin entanglement in an artificial
triangular molecule built of coherently coupled quantum dots. In a subspace of doublet states
an explicit relation of concurrence with spin correlation functions and chirality is
presented. The electric field modifies super-exchange correlations, shifts many-electron levels
(the Stark effect) as well as changes spin correlations. For some specific orientation of the
electric field one can observe monogamy, for which one of the spins is  separated from two
others. Moreover, the Stark effect manifests itself in a different spin entanglement for small
and strong electric fields. A role of magnetic flux is opposite, it leads to circulation of
spin supercurrents and spin delocalization.
\end{abstract}


\maketitle

\section{Introduction}

The last decade has seen a great interest in application of concepts from quantum information
theory, for example entanglement \cite{horodeccy}, to condensed matter theory \cite{amico}.
Since entanglement represents unique quantum correlations, the concept has been applied to
exploration of phenomena in strongly correlated many-fermion systems in order to gain insight
into the nature of quantum phase transitions. In this paper we show how entanglement is related
with a spin correlation function and how it can be controlled by an external electric and
magnetic field. We choose a system of three coherently coupled semiconducting quantum dots with
three electrons, because it can be viewed as a realization of a three qubit system, which has
recently been of great interest \cite{barenco,burkard,vinzenzo,Weinstein,yang}. In such the
system one can find two classes of truly three-partite entangled states, represented by the
Greenberger-Horne-Zeilinger (GHZ) and the Werner (W) states \cite{horodeccy,dur}. Recently
there are attempts to measure and control these states \cite{dicarlo}, as well as to apply them
in logical gates \cite{neeley}. In this paper we follow a scheme for universal quantum
computations, proposed by Di Vincenzo {\it et al}. \cite{vinzenzo} for  a spin system with
exchange interactions in quantum dots, in which logical qubits are encoded in the doublet
subspace with $S_z=+1/2$ (see also \cite{Weinstein}).

Recent experiments demonstrate that in three quantum dots one can perform coherent spin
manipulations \cite{laird10,takakura,gaudreu}. It is well known that spin manipulation can be
controlled by electric field,  for example in the systems with the spin-orbit interactions
\cite{kato,golovach}, by applying inhomogeneous static magnetic field \cite{laird,pioro}, by a
light-induced magnetic field through the dynamical Stark effect \cite{gupta} or Raman
transitions \cite{imamoglu} (see also \cite{hanson}). In our approach a role of the electric
field is different, it modifies super-exchange coupling. In the system with a triangular
geometry the electric field breaks its symmetry and changes the quantum correlations between
spins. Role of a magnetic flux is different, it induces spin supercurrents flowing around the
triangular ring and is the main decoherence source in that kind of systems \cite{choi}. One can
expect that the magnetic flux acts destructively on the entanglement.

The paper is organized as follows. In Sec. II we show that the concurrence, as a measure of
entanglement, has an explicit relation with spin correlation functions and chirality in the
triple spin system. Therefore one can have a simple interpretation of separability, monogamy
and dark spin states. Sec. III describes our system  within the Hubbard model and its canonical
transformation to the Heisenberg Hamiltonian. We show that the electric field breaks the
symmetry of the system and modifies exchange coupling, whereas the magnetic flux generates spin
chirality. Detail studies of the concurrence, as well as the spin correlation functions and
spin chirality, are presented in Sec. IV. For a special orientation of the electric field we
find a biseparable state (monogamy). Sec. V summarizes the paper.

\section{Spin correlation functions and measure of entanglement}

We begin defining wave functions for three electrons in the three qubit system. These functions
can be constructed by adding a third electron to the singlet or triplet state (see
\cite{pauncz}). Two spin subspaces are possible to define: quadruplets and doublets. The
quadruplets are the states with the quantum spin number $S=3/2$, $S_{z}=\{\pm3/2,\pm1/2\}$ and
the corresponding wave functions are constructed from a triplet state  in the form:
\begin{eqnarray}
|Q_{-3/2}\rangle = c_{3\downarrow}^\dag c_{2\downarrow}^\dag c_{1\downarrow}^\dag|0\rangle,\nonumber \\
|Q_{-1/2}\rangle = \frac{1}{\sqrt{3}}(c_{3\uparrow}^\dag c_{2\downarrow}^\dag c_{1\downarrow}^\dag + c_{3\downarrow}^\dag c_{2\uparrow}^\dag
c_{1\downarrow}^\dag + c_{3\downarrow}^\dag c_{2\downarrow}^\dag c_{1\uparrow}^\dag)|0\rangle,\nonumber \\
|Q_{1/2}\rangle = \frac{1}{\sqrt{3}}(c_{3\downarrow}^\dag c_{2\uparrow}^\dag c_{1\uparrow}^\dag + c_{3\uparrow}^\dag c_{2\downarrow}^\dag
c_{1\uparrow}^\dag + c_{3\uparrow}^\dag c_{2\uparrow}^\dag c_{1\downarrow}^\dag)|0\rangle,\nonumber \\
|Q_{3/2}\rangle = c_{3\uparrow}^\dag c_{2\uparrow}^\dag c_{1\uparrow}^\dag|0\rangle\nonumber.
\end{eqnarray}
Symbols $c_{i\sigma}^\dag$ are creation operators of an electron with the spin $\sigma$ in the
qubit \emph{i} acting on the vacuum $|0\rangle$. A linear combination
$|GHZ\rangle=\frac{1}{\sqrt{2}}(|Q_{-3/2}\rangle+|Q_{+3/2}\rangle)$ is known as the GHZ state. Two
other functions $|Q_{+1/2}\rangle$ and $|Q_{-1/2}\rangle$ are called W-states. These states are well
known in the literature \cite{dur, wang, miyake, horodeccy,amico}.

In this paper the studies are focused on the doublet subspace with the total spin $S=1/2$. These states were
proposed for exchange-interaction universal quantum computations
\cite{vinzenzo,Weinstein,yang}. In many cases, as the one considered in the next part of the
paper, the doublet is the ground state. We assume that the system is kept coherent for a time
sufficiently long in order to perform an entanglement measurement, and we fix the z-component
of the total spin $S_z=+1/2$ in further considerations. The wave function can be expressed as
\begin{eqnarray}\label{doublet}
|\Psi_{D_{1/2}}\rangle=\alpha_1|D_{1/2}\rangle_1+\alpha_2|D_{1/2}\rangle_2,
\end{eqnarray}
where:
\begin{eqnarray}\label{d1}
|D_{1/2}\rangle_1=\frac{1}{\sqrt{2}}(c_{3\downarrow}^\dag c_{2\uparrow}^\dag  - c_{3\uparrow}^\dag c_{2\downarrow}^\dag
)c_{1\uparrow}^\dag|0\rangle,\\ \label{d2}
|D_{1/2}\rangle_2=\frac{1}{\sqrt{6}}[(c_{3\downarrow}^\dag c_{2\uparrow}^\dag + c_{3\uparrow}^\dag c_{2\downarrow}^\dag) c_{1\uparrow}^\dag - 2
c_{3\uparrow}^\dag c_{2\uparrow}^\dag c_{1\downarrow}^\dag]|0\rangle,
\end{eqnarray}
In general $|\Psi_{D_{1/2}}\rangle$ should be expanded including a linear combination with the states
with double site occupation: $c_{i\uparrow}^\dag c_{j\uparrow}^\dag
c_{j\downarrow}^\dag|0\rangle$. The state $|D_{1/2}\rangle_1$ is prepared by adding third
electron to the singlet state, whereas $|D_{1/2}\rangle_2$ can be prepared from triplet or
singlet states \cite{pauncz}. This construction allows us for gaining insight into monogamy and
biseparability \cite{Acin,Sabin}.

As a measure of the entanglement we take the concurrence. In order to calculate it \cite{coffman}, we
define a reduced density matrix of a pair of electrons in the quantum dots $i$ and $j$:
$\varrho_{ij}=\text{Tr}_k\varrho$, where $i$, $j$ and $k$ denote different quantum dots, and
$\varrho$ is a density matrix $\varrho=|\Psi_D\rangle \langle \Psi_D|$ for the doublet
subspace. Next we derive a matrix:
\begin{eqnarray}
R_{ij}=\varrho_{ij}(\sigma_y\otimes\sigma_y)\varrho^{\ast}_{ij}(\sigma_y\otimes\sigma_y),
\end{eqnarray}
where $\sigma_y$ is a Pauli matrix and the asterisk denotes complex conjugation
of $\varrho_{ij}$. The concurrence is calculated as:
\begin{eqnarray}\label{con}
C_{ij}=\max\{0,\lambda_1-\lambda_2-\lambda_3-\lambda_4\},
\end{eqnarray}
where $\lambda_1$, $\lambda_2$, $\lambda_3$, $\lambda_4$ are square roots of eigenvalues of
$R_{ij}$ in descending order. $C_{ij}$ can take values between zero (for separate states) and
one (for fully quantumly entanglement states). Using this definition one can calculate the
concurrence for the doublet representation (\ref{doublet}):
\begin{eqnarray}\label{c12}
C_{12}=\frac{2}{3} |\alpha_2||\sqrt{3}\alpha_1+\alpha_2|,\\
\label{c13}
C_{13}=\frac{2}{3} |\alpha_2||\sqrt{3}\alpha_1-\alpha_2|,\\
\label{c23}
C_{23}=\frac{1}{3} |3\alpha_1^2-\alpha_2^2|.
\end{eqnarray}

Let us now calculate spin correlation functions in the doublet subspace
(\ref{doublet}):
\begin{eqnarray}\label{s1s2}
\langle \mathbf{S}_1\cdot\mathbf{S}_2\rangle=\frac{1}{4}[-\sqrt{3}(\alpha_1 \alpha_2^*+\alpha_1^* \alpha_2)-2|\alpha_
2|^2],\\ \label{s1s3}
\langle \mathbf{S}_1\cdot\mathbf{S}_3\rangle=\frac{1}{4}[\sqrt{3}(\alpha_1 \alpha_2^*+\alpha_1^* \alpha_2)-2|\alpha_2|^2],\\ \label{s2s3}
\langle \mathbf{S}_2\cdot\mathbf{S}_3\rangle=\frac{1}{4}(-3|\alpha_1|^2+|\alpha_2|^2).
\end{eqnarray}
In general the coefficients $\alpha_1$ and $\alpha_2$ can be complex, for example
in the presence of the magnetic flux. We show later that in this case a spin
supercurrent occurs with a nonzero value of chirality
\begin{eqnarray}
\langle \mathbf{S}_1\cdot(\mathbf{S}_2\times \mathbf{S}_3)\rangle=\text{i}\frac{\sqrt{3}}{4}(\alpha_1 \alpha_2^*-\alpha_1^* \alpha_2).\label{cij=sisj}
\end{eqnarray}

 Comparing these results with the concurrence $C_{ij}$
(\ref{c12})-(\ref{c23}) one can find, after some algebra, the following relation:
\begin{eqnarray}
C_{ij}=\frac{4}{3}|\langle \mathbf{S}_i\cdot\mathbf{S}_j+\text{i}\;\mathbf{S}_1\cdot(\mathbf{S}_2\times \mathbf{S}_3)\rangle|.\label{cij=sisj}
\end{eqnarray}
This fact allows us to propose the expectation value of spin correlation
functions  and chirality as an alternative measure of the entanglement.

 In a multi-qubit system it is interesting to
define monogamy of entanglement. If two qubits are fully quantumly entanglement then they
cannot be correlated with the third one \cite{bennett}. In this case one says about monogamy, which
expresses the nonshareability of entanglement. For a three qubit system monogamy can be
measured by the concurrence $C_{i(jk)}$ between a qubit \textit{i} and other qubits $j,k$,
which is given by the one-tangle $C^2_{i(jk)}=4 \det\varrho_i$, where
$\varrho_i=\text{Tr}_{jk}\varrho$. For biseparable states $C^2_{i(jk)}=0$, which means that the
qubits $j$ and $k$ are fully quantumly entanglement whereas the qubit $i$ is separated. This
quantity is related with the linear entropy $S_L(\varrho_i)\equiv 2[1-\text{Tr}(\varrho^2_i)]=4
\det\varrho_i$. Monogamy of entanglement satisfies the relation
\begin{eqnarray}
C_{i(jk)}^2 \geq C_{ij}^2+C_{ik}^2
\end{eqnarray}
proved by Coffman, Kundu and Wooters \cite{coffman}. For our case, in the doublet subspace, we
have the equality $C_{i(jk)}^2 = C_{ij}^2+C_{ik}^2$, as one could expect for pure states.

\section{Model of correlated spins on a triangular system of quantum dots}

In this section we would like to present a specific three qubit example, namely an artificial
triangular molecule built of coherently coupled semiconducting quantum dots. We show first that
an external electric field and a magnetic flux can modify spin states, and later, in the next
section, how these external fields influence the entanglement between spins.

\begin{figure}[ht]
\includegraphics[width=0.25\textwidth, clip]{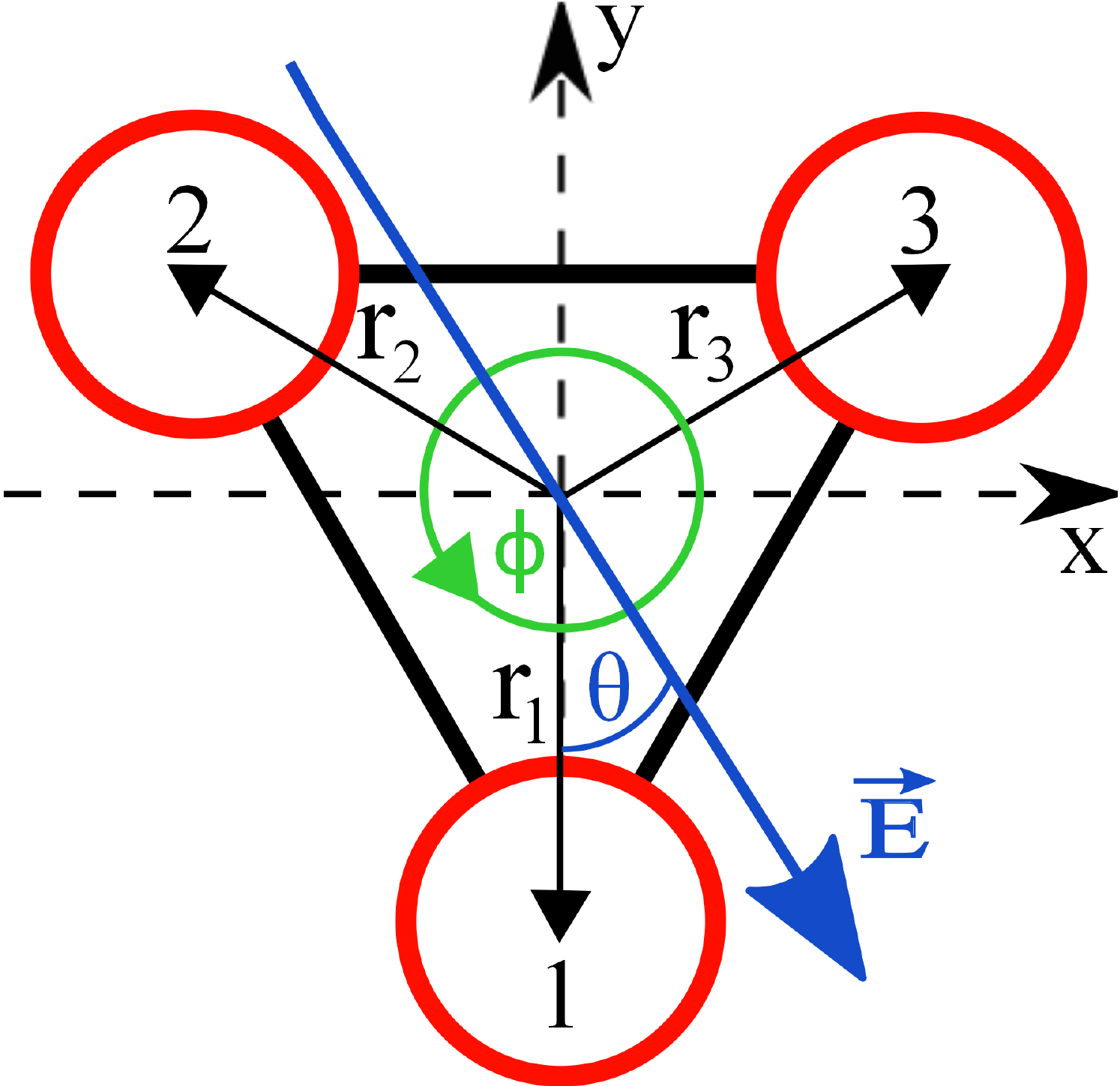}
\caption{Model of a system of Three Quantum Dots placed in an external electric filed $\mathbf{E}$ and a magnetic flux $\Phi$.  }
\label{model}
\end{figure}

Our system of three quantum dots (Fig.\ref{model}) is described by the Hubbard model
\begin{equation}
\begin{split}
\hat{H}&= \sum_{i,\sigma}\{\epsilon_i + Eer\cos[\theta+(i-1)2\pi/3]\}n_{i\sigma}\\
&+t \sum_{i\neq j,\sigma}(e^{i\phi/3}c_{i\sigma}^\dag c_{j\sigma}+h.c.)
 +U \sum_i n_{i\downarrow}n_{i\uparrow}.
\label{hubbard}
\end{split}
\end{equation}
Here, $\widetilde{\epsilon}_i=\epsilon_i+Eer\cos[\theta+(i-1)2\pi/3]$ corresponds to
a shift of a local single electron level $\epsilon_i$ in the electric field
$\mathbf{E}$. The polarization energy $\mathbf{E}\cdot\mathbf{P}=e\sum_{i\sigma}
\mathbf{E}\cdot \mathbf{r_i}n_{i\sigma}=Eer
\sum_{i\sigma}\cos{[\theta+(i-1)2\pi/3]}n_{i\sigma}$, where: \emph{e} - the
electron charge, $\mathbf{r}_i$ denotes a vector of the \emph{i}-qubit position and
$\theta$ is an angle between $\mathbf{r}_1$ and $\mathbf{E}$, $n_{i\sigma}$ -
an electron number operator.  Later for simplicity we put $\epsilon_i=0$ and denote
$g_E=Eer$. The second term in (\ref{hubbard}) describes electron hopping between
nearest quantum dots in the presence of the magnetic flux $\Phi$ enclosed in the
triangle. According to the Peierls scaling an electron gains during hopping a
phase shift $\phi= 2 \pi \Phi/(hc/e)$. The Coulomb onsite interaction of
electrons on the quantum dots is included in the last term.

Using this model we can calculate all properties of the system numerically. In
particular for three electrons we calculate the spin correlation functions as
well as the concurrence and influence of the electric field as well as the magnetic flux.
To understand the results we use a canonical transformation \cite{bulka1}
of the Hubbard Hamiltonian (\ref{hubbard}) to an effective Heisenberg
Hamiltonian. Taking the hopping integral $t$ and the electric field $g_E$ as
small parameters with respect to the Coulomb interaction $U$, one can get the
effective Heisenberg Hamiltonian \cite{kostyrko,dassarma}:
\begin{eqnarray}\label{heisenberg}
\hat{H}_{\text{eff}} = \sum_{i<j}J_{ij}
(\mathbf{S}_i\cdot\mathbf{S}_j- \frac{1}{4})+J_{\chi}\;\mathbf{S}_1\cdot(\mathbf{S}_2\times \mathbf{S}_3) .
\end{eqnarray}
The first term describes the superexchange coupling between spins, for which the
exchange parameter $J_{ij}$ can be calculated to the third order in $t/U$
\cite{kostyrko}
\begin{eqnarray}\label{heisenbergJ}
J_{ij}&=&2|t_{ij}|^2(\Delta_{ji}^{-1}+\Delta_{ji}^{-1})+
\left(t_{ji}t_{ik}t_{kj}+t_{ji}^*t_{ik}^*t_{kj}^*\right)\times
  \nonumber \\
  &&(\Delta^{-1}_{ij}\Delta^{-1}_{kj}  +
  \Delta^{-1}_{ji}\Delta^{-1}_{ki}
  +  \Delta^{-1}_{ki}\Delta^{-1}_{kj} \nonumber\\
  &&-\Delta^{-1}_{ji}\Delta^{-1}_{jk}
  - \Delta^{-1}_{ij}\Delta^{-1}_{ik}
  - \Delta^{-1}_{ik}\Delta^{-1}_{jk})\,,
\end{eqnarray}
where $\Delta_{ij}=(U + \widetilde{\epsilon}_j - \widetilde{\epsilon}_i)$. In the
limit $g_E\ll U$ one can get an explicit form of $J_{ij}$:
\begin{eqnarray}
J_{12}=\frac{4t^2}{U}+\frac{3t^2g_E^2(2+\cos{2\theta}+\sqrt{3}\sin{2\theta})}{U^3}\nonumber\\
-\frac{12t^3\cos\phi\; g_E(\cos{\theta}-\sqrt{3}\sin{\theta})}{U^3}, \label{j12}
\end{eqnarray}
\begin{eqnarray}
J_{13}=\frac{4t^2}{U}+\frac{3t^2g_E^2(2+\cos{2\theta}-\sqrt{3}\sin{2\theta})}{U^3}\nonumber\\
-\frac{12t^3\cos\phi\; g_E(\cos{\theta}+\sqrt{3}\sin{\theta})}{U^3}, \label{j13}
\end{eqnarray}
\begin{eqnarray}
J_{23}=\frac{4t^2}{U}+\frac{6t^2g_E^2 (1-\cos{2\theta})}{U^3}+\frac{24t^3 \cos\phi\; g_E\cos{\theta}}{U^3}. \label{j23}
\end{eqnarray}
One can see that the second term is proportional to $g_E^2$ and corresponds to
the quadratic Stark effect. When the electric field rotates, this term leads to
oscillations with the period $\pi$. The linear Stark effect is described by the
third term, which corresponds to the period of oscillations equal to $2\pi$. The
linear term in $J_{23}$ is always positive, whereas for $J_{12}$ and $J_{13}$ the
linear terms can be negative. At $\theta=0$ one can see that $J_{23}$ increases
linearly with $g_E$, whereas the couplings $J_{12}=J_{13}$ and they first decrease, and next
increase quadratically for a larger $g_E$. At $g_E=4 t \cos \phi$ we get
$J_{12}=J_{13}=J_{23}$ and the system becomes uniform once again.

The second term in the effective Hamiltonian (\ref{heisenberg}) describes
chirality of  electrons in the presence of the magnetic flux. The term is
connected with the Aharonov-Bohm effect and with the persistent currents moving around
the flux enclosed by the three quantum dot ring. The coupling parameter calculated to
the third order in $t/U$ is given by \cite{kostyrko}
\begin{eqnarray}\label{Jchi}
  J_{\chi} = -\textsl{i}
(t_{ji}t_{im}t_{mj}-t_{ji}^*t_{im}^*t_{mj}^*)
  (\Delta^{-1}_{ij}\Delta^{-1}_{mj} +\Delta^{-1}_{ji}\Delta^{-1}_{jm}\nonumber\\
  + \Delta^{-1}_{ji}\Delta^{-1}_{mi}
  + \Delta^{-1}_{im}\Delta^{-1}_{ij} + \Delta^{-1}_{im}\Delta^{-1}_{jm}
  +   \Delta^{-1}_{mi}\Delta^{-1}_{mj})\,,
\end{eqnarray}
which in the limit $g_E\ll U$ simplifies to the form $J_{\chi}=-12t^3\sin
\phi/U^2$. This parameter depends on the electric field in higher order terms of
the expansion, but we neglect them in our studies.

Using the Heisenberg model (\ref{heisenberg}) one can derive many physical quantities
analytically, in particular energy for the quadruplet and doublet states
presented in the second chapter. For $|GHZ\rangle$ and $|W\rangle$ state one has the energy:
\begin{eqnarray}
E_{GHZ}=E_W=0.
\end{eqnarray}
These states are independent of the electric field because the spins cannot be
transferred between quantum dots (due to the Pauli exclusion principle). In the
doublet subspace $[|D_{1/2}\rangle_1, |D_{1/2}\rangle_2]^T$  the effective Hamiltonian is expressed as
\begin{eqnarray}
H_{\text{eff}}=\left[\begin{array}{cc}
-\frac{3}{4}(J_{12}+4J_{23}+J_{31})& \frac{\sqrt{3}}{4}(J_{31}-J_{12}+\text{i} J_{\chi}) \\
\frac{\sqrt{3}}{4}(J_{31}-J_{12}-\text{i} J_{\chi})  & -\frac{3}{4}(J_{12}+J_{31})
\end{array}
\right].\nonumber\\
\end{eqnarray}
The eigenenergies are
\begin{equation}
E_{D}^{\pm}=-\frac{1}{2}(J_{12}+J_{23}+J_{13})\pm\frac{\Delta}{2},
\end{equation}
where $\Delta=[3
J_{\chi}^2/4+J_{12}^2+J_{23}^2+J_{13}^2-J_{12}J_{13}-J_{12}J_{23}-J_{13}J_{23}]^{1/2}$.
For the considered case the doublet states $E_{D}^\pm$ are below $E_{GHZ}$ and
$E_W$. The corresponding eigenfunctions may be written as
\begin{eqnarray}\label{doubletH}
|\Psi_{D}\rangle^\pm=\frac{z^{\pm}}{\sqrt{1+|z^{\pm}|^2}}\;|D_{1/2}\rangle_{1} +\frac{1}{\sqrt{1+|z^{\pm}|^2}}\;|D_{1/2}\rangle_{2},
 \label{fun}
\end{eqnarray}
where $z^{\pm}=(J_{12}-2J_{23}+J_{13}\pm 2\Delta)/[\sqrt{3}(J_{13}-J_{12}-\text{i}J_{\chi})]$.

\section{Influence of external fields on entanglement}

We calculated the spin-spin correlation functions $\langle
\mathbf{S}_i\cdot\mathbf{S}_j\rangle$, the concurrence $C_{ij}$ and $C_{i(jk)}$ for the Hubbard
model (\ref{hubbard}) as well as for the effective Heisenberg model (\ref{heisenberg}). The
results for both the models were the same, within the relative accuracy better than 0.01, for
the parameters used below. Therefore our analysis is performed for the Heisenberg model
(\ref{heisenberg}), which is much simpler.

\subsection{Role of electric field}

\begin{figure}[ht]
\centering
\includegraphics[width=0.48\textwidth, clip]{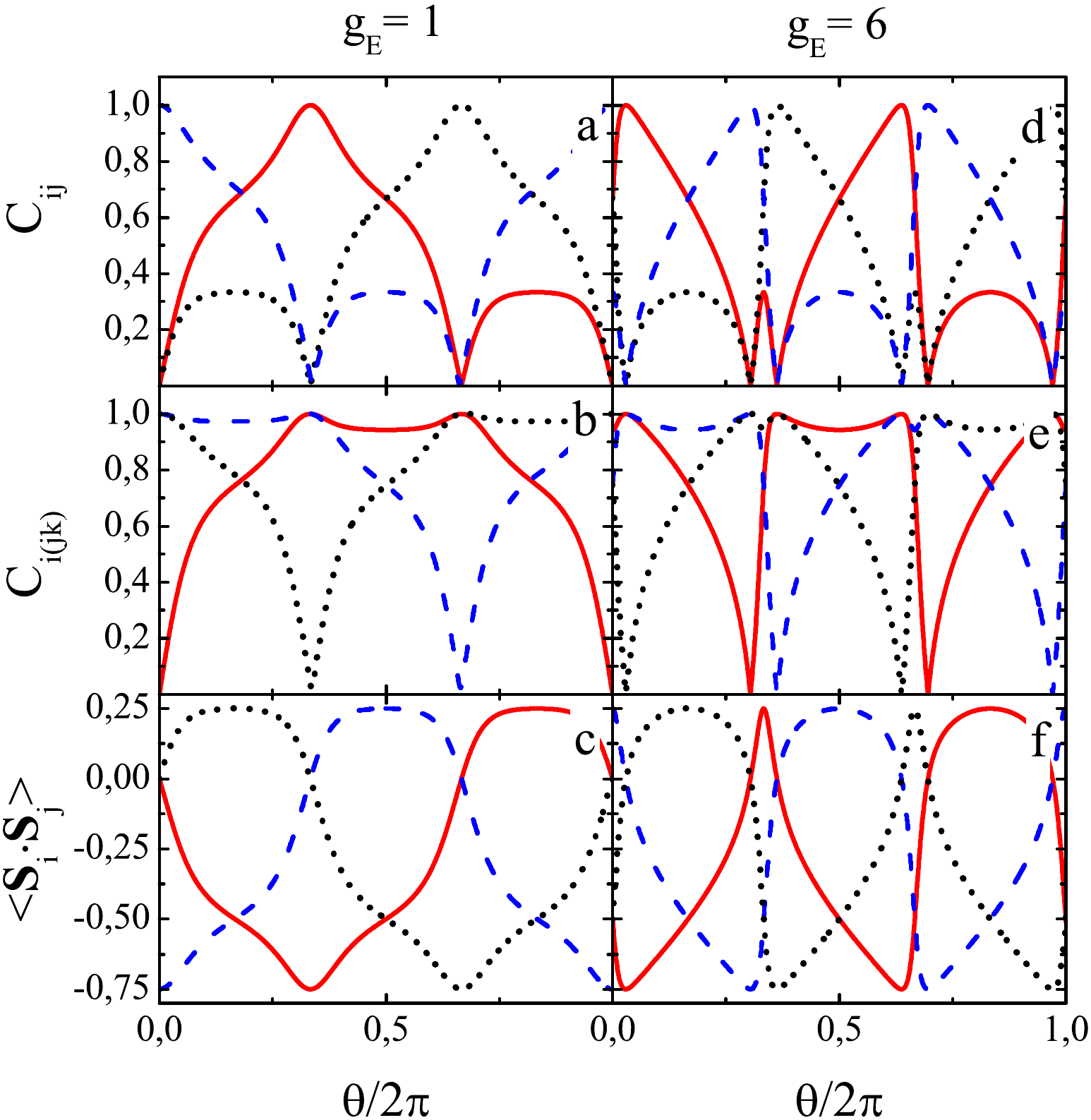}
\caption{(color on-line) Dependence of the concurrence $C_{ij}$, $C_{i(jk)}$ and the spin correlation functions $\langle
\mathbf{S}_i\cdot\mathbf{S}_j\rangle$ vs. the angle $\theta$ of the electric field
calculated for the ground state.  The (red) solid curves represent: $C_{12}$, $C_{1(23)}$ and $\langle
\mathbf{S}_1\cdot\mathbf{S}_2\rangle$; the (blue) dashed curves: $C_{23}$, $C_{2(13)}$
and $\langle \mathbf{S}_2\cdot\mathbf{S}_3\rangle$; the (black) dotted curves: $C_{13}$,
$C_{3(12)}$ and $\langle \mathbf{S}_1\cdot\mathbf{S}_3\rangle$.
Left panel is plotted for a small electric field $g_E=1$, right one is for a large electric field  $g_E=6$. The
other parameters are: $U=20$, $\epsilon_i=0$ and $t$ is taken as unity.}
\label{fig2}
\end{figure}

Let us first study how entanglement can be modified by an electric field only (in the absence
of the magnetic field, i.e. for $\phi=0$ and $J_{\chi}=0$). The electric field breaks the
symmetry of the system, induces polarization (altought its value is small due to the strong onsite
repulsion $U$). The electric field also changes the concurrence $C_{ij}$,
$C_{i(jk)}$ and the spin correlation function $\langle \mathbf{S}_i\cdot\mathbf{S}_j\rangle$.
Fig.\ref{fig2} shows these quantities for the ground state as a function of the angle $\theta$
of the electric field with respect to the axes of the system. The left panels are plotted for a small electric filed ($g_E=1$), when
the linear Stark effect dominates. The period of oscillation is $2\pi$ for $C_{ij}$,
$C_{i(jk)}$ and $\langle \mathbf{S}_i\cdot\mathbf{S}_j\rangle$. One can see in Fig.\ref{fig2}a
that for $\theta=0$ the concurrence $C_{12}=C_{13}=0$ and $C_{23}=1$. The spins in the qubit
$2$ and $3$ are fully entangled which means monogamy. The spin in the qubit 1 is separated
from two others (see also the red solid curve Fig.2b showing that $C_{1(23)}=0$ at $\theta=0$). The spin
correlation functions (Fig.\ref{fig2}c) $\langle \mathbf{S}_1\cdot\mathbf{S}_2\rangle=\langle
\mathbf{S}_1\cdot\mathbf{S}_3\rangle=0$ and $\langle
\mathbf{S}_2\cdot\mathbf{S}_3\rangle=-0.75$ which means that the spins $2$ and $3$ are in the
singlet state. The wave function has the form: $|\Psi_{D_{1/2}}\rangle=|D_{1/2}\rangle_1$.
According to the classification \cite{Sabin} we have the simply biseparable states for which the
density matrix can be written as $\rho=\rho_1\otimes \rho_{23}$. At $\theta=0$ the electric
field is oriented to the quantum dot $1$ and the system has the mirror symmetry with the
exchange couplings $J_{12}=J_{13} < J_{23}$ [see (\ref{j12})-(\ref{j23})]. A similar situation
occurs for $\theta=2\pi/3$ and $4\pi/3$, only if $g_E<4|t|$.

For a large electric field $g_E>4|t|$ the situation is different and more complex (see the
right panels in Fig.\ref{fig2}). Now, the period of oscillations is changed due to the
quadratic Stark effect [the second term in Eqs.(\ref{j12})-(\ref{j23})].  For $\theta=0$ the
ground state is $|D_{1/2}\rangle_2$ and we do not observe monogamy. The spin in the qubit 1 is
separable from two others for $\theta_0$ close to $2\pi/3$ and $4\pi/3$ (see $C_{1(23)}$ in
Fig.\ref{fig2}e). The condition for monogamy is the mirror symmetry of the system. Using
Eqs.(\ref{j12})-(\ref{j23}) and $J_{12}=J_{13} < J_{23}$ one can get the angle
$\theta_0=\arccos(-2t/g_E)$ for separability of the qubit 1. Fig.\ref{fig2}f shows that at
$\theta_0$ the quantum dots 2 and 3 are fully quantumly entanglement with $\langle
\mathbf{S}_2\cdot\mathbf{S}_3\rangle=-3/4$ (see the dashed curve).

\subsection{Role of magnetic flux}

In the presence of the magnetic flux ($\phi\neq 0$) chirality of the spin system  [described
by the last term in (\ref{heisenberg})] becomes relevant
\cite{dassarma,Wen,Katsura,Motrunich,Bulaevskii,trif,Hsieh}. Recently Hsieh {\it et al.}
\cite{Hsieh} proposed to use chirality for quantum computations. Their quantum circuits are
based on qubits encoded in chirality of electron spin complexes in systems of triangular
quantum dot molecules. The magnetic flux removes degeneracy between the states with different
orbital momenta and leads to spin supercurrents circulating around the triangle
\cite{Wen,Katsura}. We take the expectation value of the operator \cite{trif}
\begin{eqnarray}\label{chiralop}
C_z=(4/\sqrt{3})\mathbf{S}_1\cdot(\mathbf{S}_2\times \mathbf{S}_3)
\end{eqnarray}
as a measure of chirality.\cite{Wen} In the basis of the doublets (\ref{doublet}) and using
(\ref{doubletH}) one gets the expectation value
\begin{eqnarray}\label{chiralop}
\langle C_z \rangle=\text{i} (\alpha_1 \alpha^*_2-\alpha^*_1 \alpha_2)=-\frac{\sqrt{3}J_{\chi}}{2\Delta}
\end{eqnarray}
in the ground state. It means that chirality depends on the splitting $\Delta$ between the
doublets which can be controlled by the electric field as well.

In the absence of the electric field (when
all exchange couplings are equal $J_{ij}=J$) the expectation value $\langle
C_z\rangle=\pm 1$, because the supercurrent is $\propto \sin \phi$ and
circulates clockwise or anticlockwise (for $0<\phi<\pi$ or
$\pi<\phi<2\pi$, respectively). For this case the spins are delocalized and
the expectation values of the spin correlation functions $\langle
\vec{S}_i\cdot\vec{S}_j\rangle=-1/4$. The concurrence $C_{ij}=1/3$ and
$C_{i(jk)}=\sqrt{2}/3$, which describes maximal spin mixing.

\begin{figure}[ht]
\centering
\includegraphics[width=0.4\textwidth, clip]{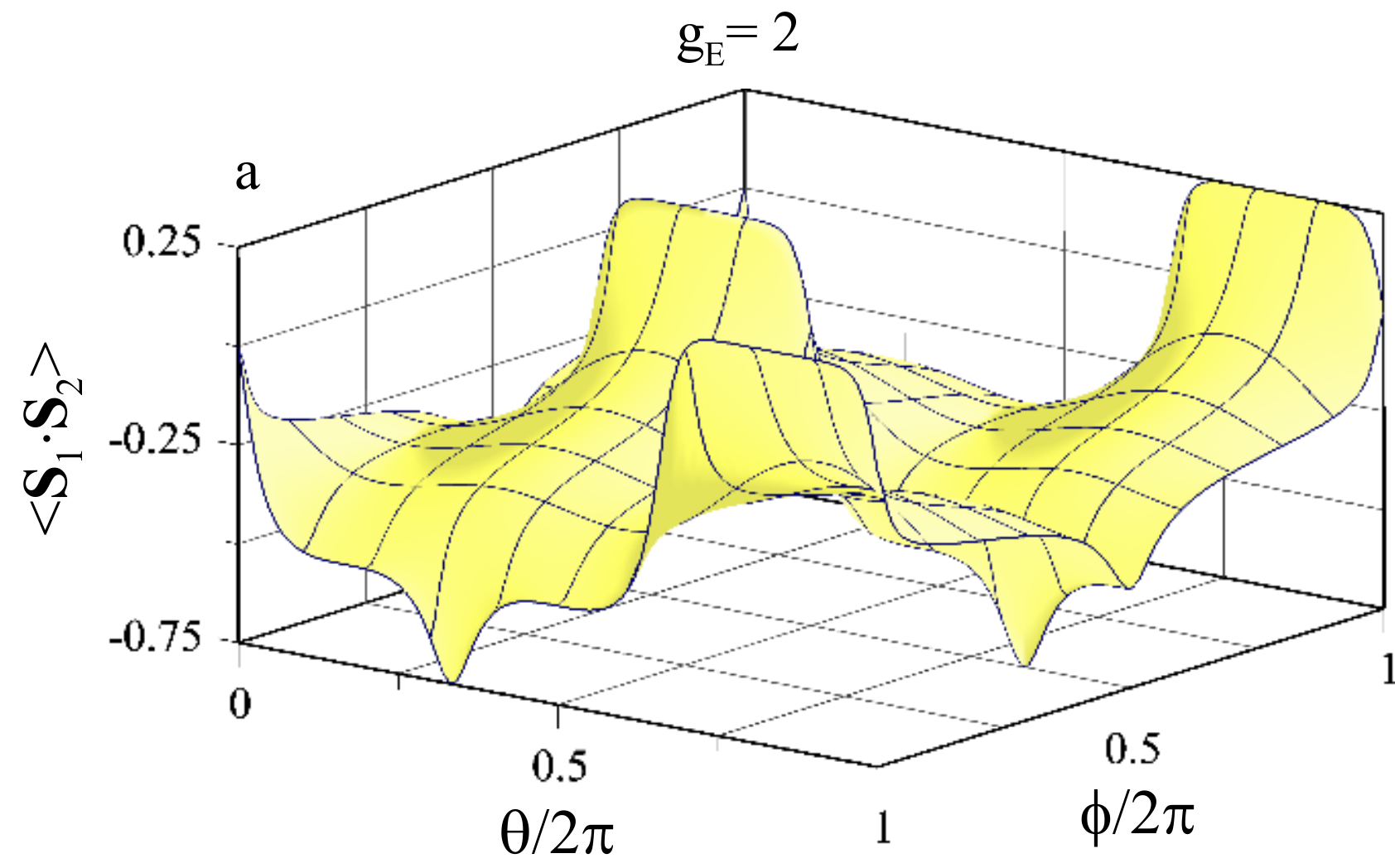}\\
\includegraphics[width=0.4\textwidth, clip]{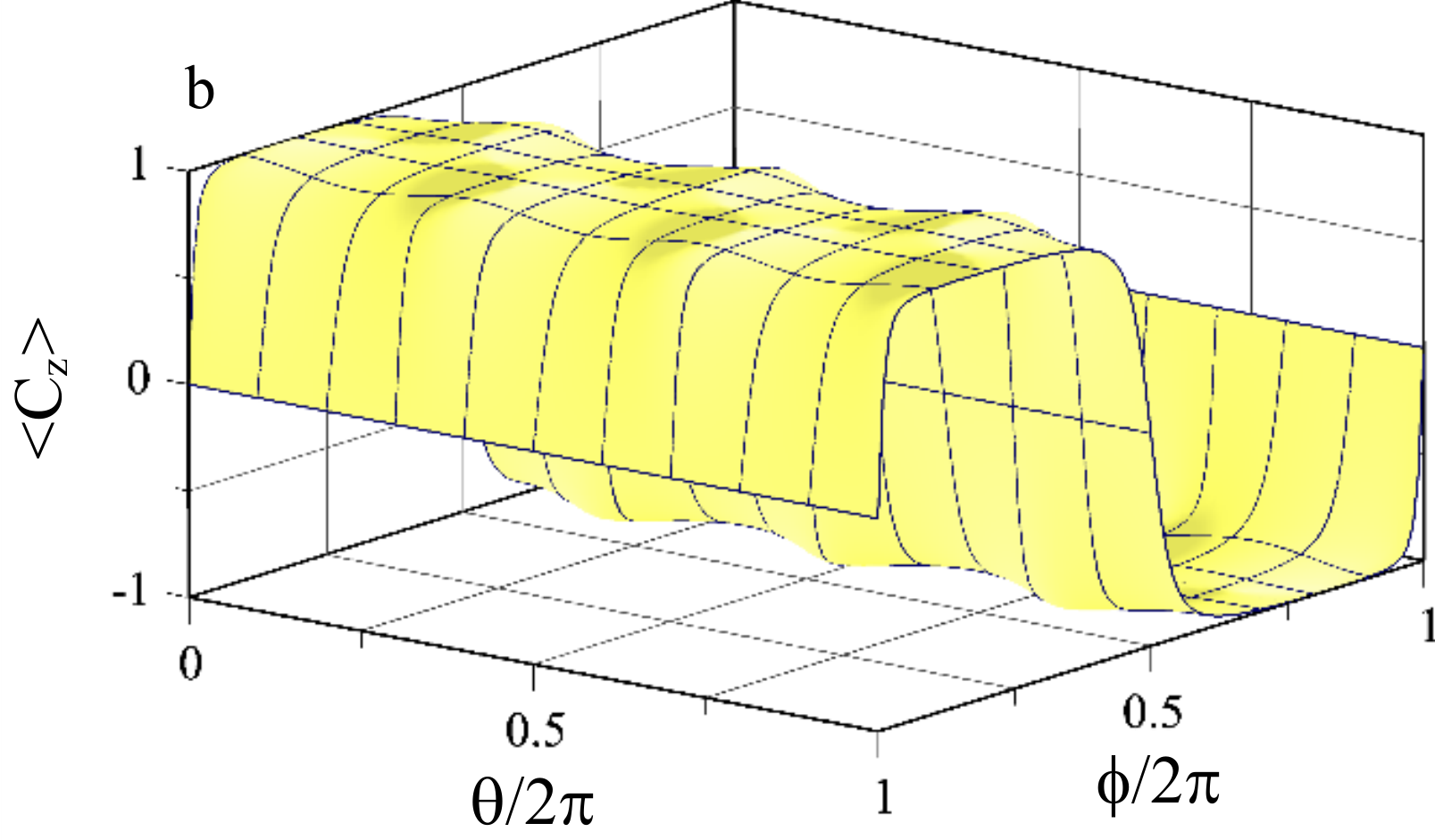}\\
\includegraphics[width=0.4\textwidth, clip]{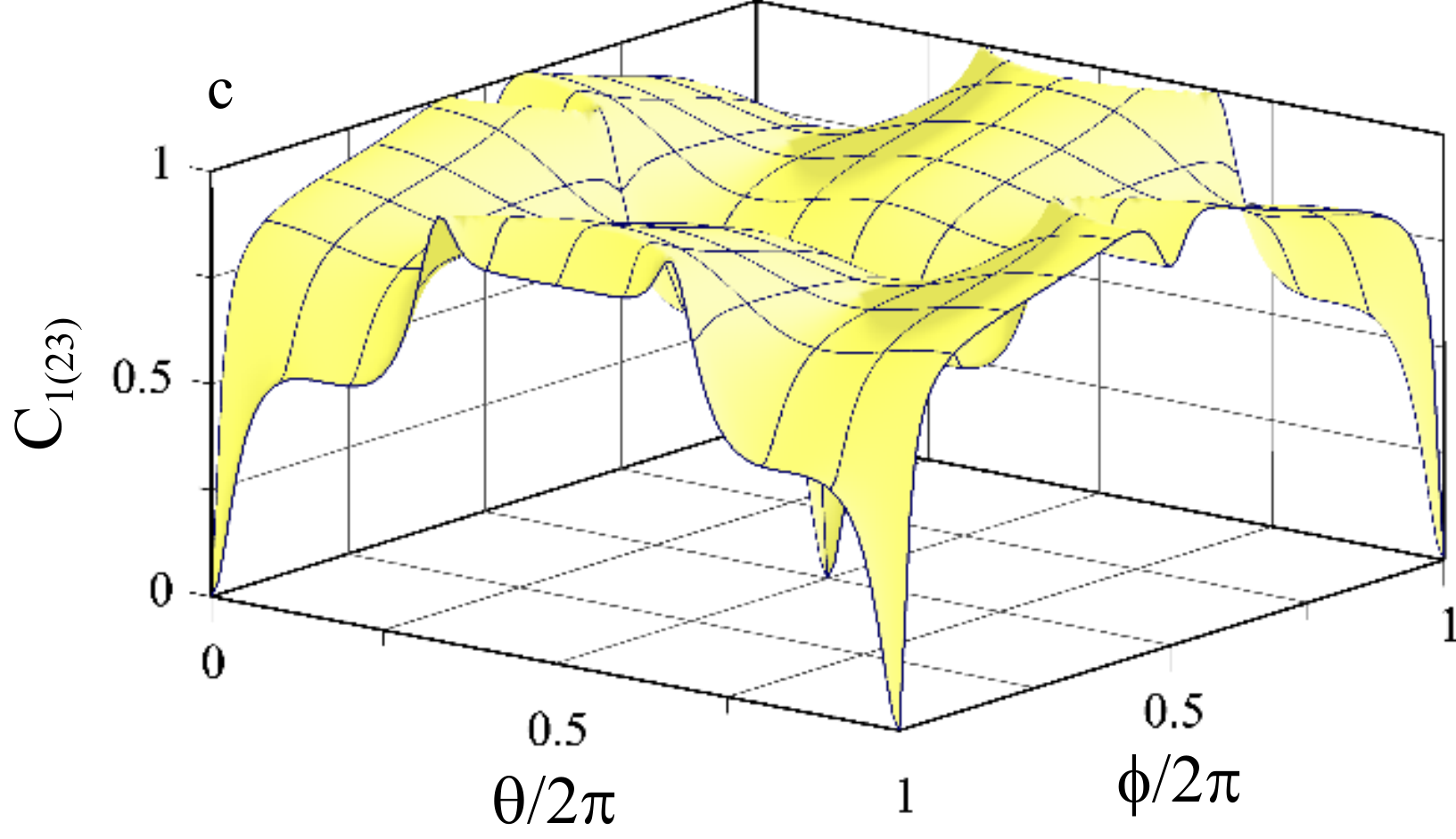}
\caption{Plots of the spin correlation function $\langle \mathbf{S}_1\cdot\mathbf{S}_2\rangle$,
the chirality $\langle C_z\rangle$  and the concurrence $C_{1(23)}$ in the plane $\theta$ and
$\phi= 2 \pi \Phi/(hc/e)$ for a small electric field $g_E=2$ and the other parameters the same as in Fig.2.}
\end{figure}

\begin{figure}[ht]
\centering
\includegraphics[width=0.4\textwidth, clip]{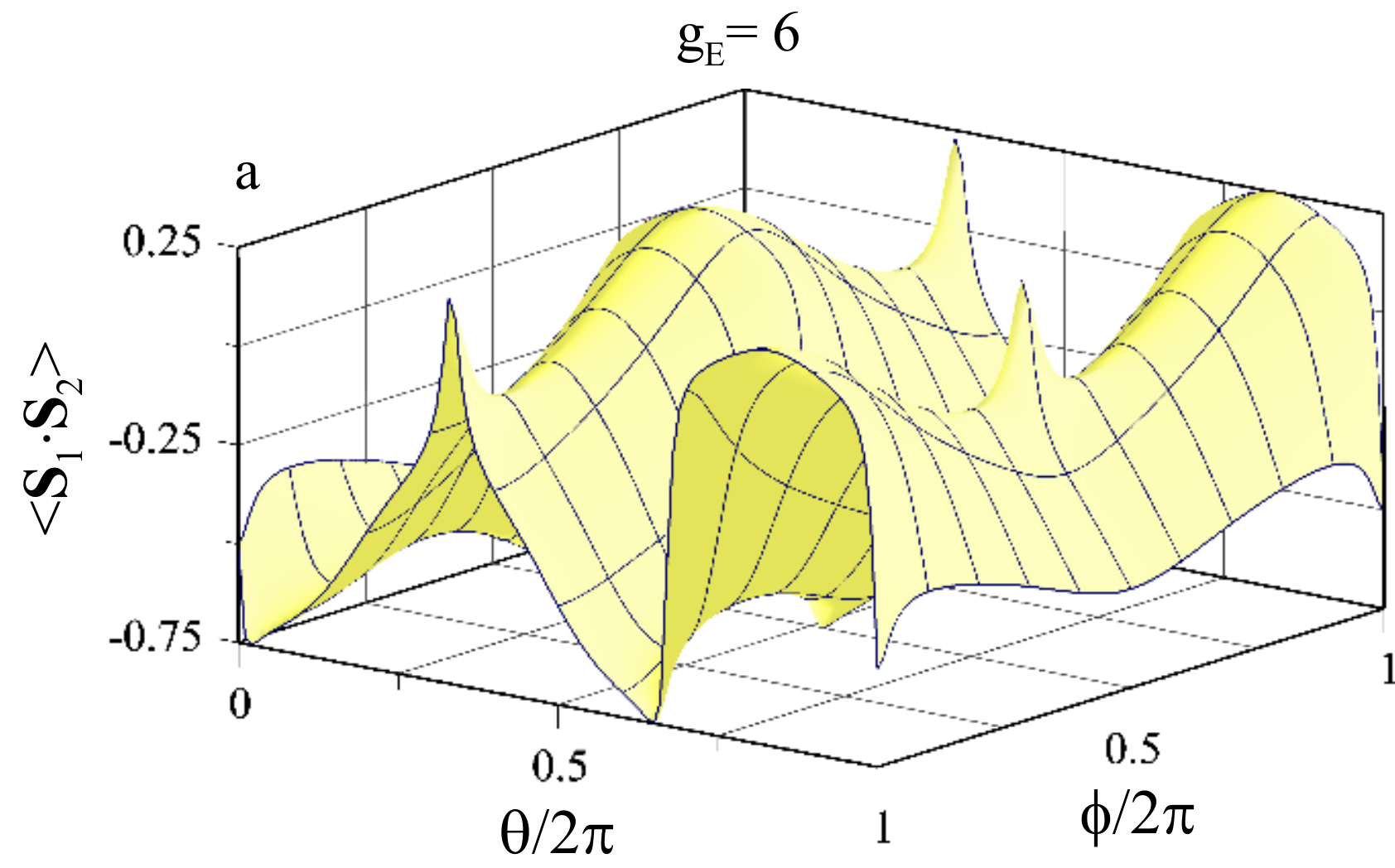}\\
\includegraphics[width=0.4\textwidth, clip]{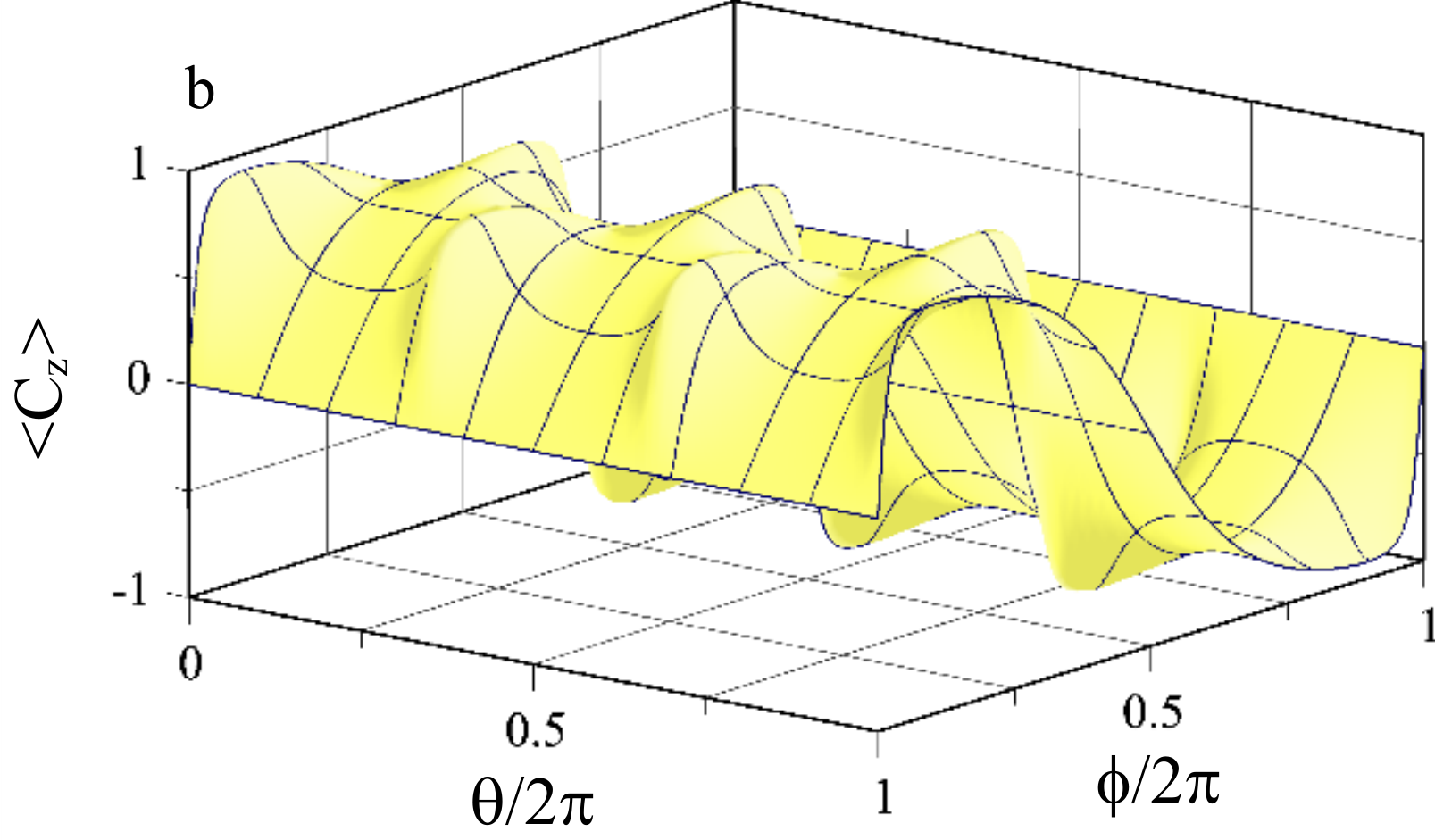}\\
\includegraphics[width=0.4\textwidth, clip]{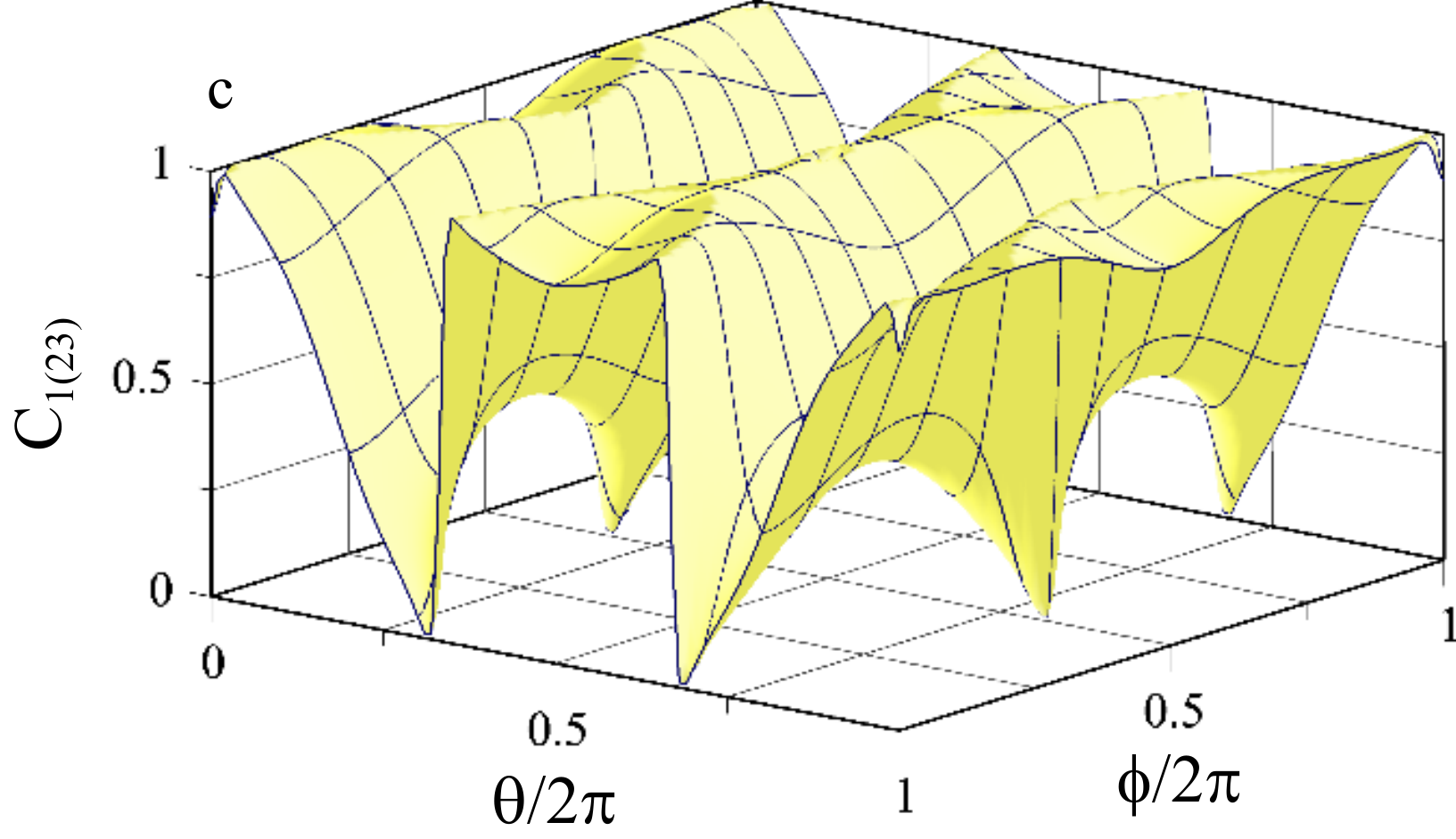}
\caption{Plots of $\langle \mathbf{S}_1\cdot\mathbf{S}_2\rangle$,
$\langle C_z\rangle$ and $C_{1(23)}$ in the $\theta$-$\phi$ plane for a large electric field $g_E=6$.}
\end{figure}
In a general situation, in the presence of both the fields, one can expect a competition between
localization and delocalization of spins. These processes should manifest themselves in
the spin correlation function $\langle \mathbf{S}_i\cdot\mathbf{S}_j\rangle$ (in $C_{ij}$,
$C_{i(jk)}$) as well as in the expectation value of chirality $\langle C_z\rangle$. Fig.3 and 4
present plots for $\langle \mathbf{S}_1\cdot\mathbf{S}_2\rangle$, $\langle C_z\rangle$ and
$C_{1(23)}$ in the $\theta$-$\phi$ plane for a small and large electric field, respectively.
Close to $\phi=0$  and $\pi$ the spin supercurrent ($\propto \sin \phi$) is small and the Stark
effect dominates. In the other regions the magnetic flux becomes relevant and $\langle
C_z\rangle$ reaches its extremal value $\pm1$ (see the plateau with small oscillations caused
by the electric field in Fig.3b). According to (\ref{chiralop}) $\langle C_z\rangle$ is
inversely proportional to the energy gap $\Delta$ between the doublet states and reaches
maximal values at the symmetry points $\theta=0$, $2\pi/3$ and $4\pi/3$ (see also Fig.4b). The
functions $\langle \mathbf{S}_1\cdot\mathbf{S}_2\rangle$ and $C_{1(23)}$  are very sensitive to
symmetry breaking caused by the electric field. They show changes when the electric field
becomes larger $g_E >4|t|\cos \phi$ (compare Fig.3 and 4). A detail analysis of the plots shows
two contributions to the Stark effect: a linear and quadratic ones. The magnetic flux reduces
the linear component, and the quadratic Stark effect dominates [see also
Eqs.(\ref{j12})-(\ref{j23}) for the exchange couplings $J_{ij}$]. If $g_E$ increases, the spins
become more localized and the amplitude of the Stark oscillations, seen in $\langle
\mathbf{S}_1\cdot\mathbf{S}_2\rangle$ and $C_{1(23)}$, increases. The localization process is
monotonic, we could not observe any drastic changes - in contrast to the situation in the
region close to $\phi=0$ and $\pi$ when the spin correlation functions and the concurrence
change drastically their characteristics in large electric fields.

\section{Conclusions}
Summarizing, we showed that entanglement can be controlled by the electric field in three spin
system of three coherently coupled quantum dots with a triangular geometry. The studies were
focussed on bipartite entanglement in the subspace of doublets with $S_z=+1/2$, for which the
concurrence was related to the spin correlation function and the spin chirality,
Eq.(\ref{cij=sisj}). This relation was exemplified for the Hubbard model and its canonical
transformation to the effective Heisenberg model. The super-exchange coupling exhibits a linear
and quadratic dependence on the electric field (the linear and quadratic spin Stark effect),
which manifests itself in different periods of oscillations of the concurrence and the spin
correlation functions when the electric field changes its orientation. The competition between
these two Stark effects leads to different characteristics of the concurrence for a small and
large electric field. For a special field orientation we found a biseparable state, for which
one of the spins is separated from two others fully quantumly entangled (monogamy). For
small fields one should direct the field precisely toward the quantum dot ($\theta_0=0$,
$2\pi/3$ and $4\pi/3$) to get the spin separation at this dot. In the case of the large electric
field its orientation $\theta_0$ should be different: the angle $\theta_0$ depends on the
strength of the field and it should be directed toward one of the opposite quantum dots.

We also considered a role of spin chirality on entanglement. The magnetic flux $\Phi$ induces
circulation of spin supercurrents and leads to spin delocalization. The bipartite concurrence
becomes uniform. For small electric fields the spins are delocalized, the Stark effect can be
only seen in a very narrow range of $\Phi$. For larger fields the Stark effect becomes more
visible, the concurrence and the spin correlations exhibit oscillations as a function of the
angle $\theta$ of the electric field. Analyzing the oscillations one can see how the magnetic flux
modifies relative contributions of the linear and quadratic Stark effect to entanglement.

Our studies can be also related to recent experiments on coherent spin manipulation in three
quantum dots \cite{laird10,takakura,gaudreu}. We showed that the scheme proposed by Di Vincenzo
{\it et al}. \cite{vinzenzo}, with logical qubits encoded in the doublet subspace
$|D_{1/2}\rangle_1$ and $|D_{1/2}\rangle_2$, can be realized due to the spin Stark effect, in
which the electric field changes spin entanglement. The ground state is a superposition of the
doublet states, which can be controlled changing orientation of the electric field. Therefore,
this effect can be used to preparation of a proper initial quantum state and control a logical
operation. Let us point out that Di Vincenzo's {\it et al}.  scheme is different than
operations between quadruplet and doublet states performed by Gaudreau {\it et al}.
\cite{gaudreu} in the experiment in triple quantum dots and between the singlet and triplet
states in double quantum dots \cite{2qd}, where passages were accompanied by reorientation of
nuclear spins in quantum dots. Within Di Vincenzo's {\it et al}.  scheme the dynamical passages
are performed only between the doublet states (without nuclear spins) and total spin $S=1/2$ as
well as its z-component $S_z=1/2$ are conserved.

\acknowledgments{ We would like to thank Anton Ram\v{s}ak and Guido Burkard for stimulating
discussions. This work was supported by Ministry of Science and Higher Education (Poland) from
sources for science in the years 2009-2012 and by the EU project Marie Curie ITN NanoCTM.}


\begin{thebibliography}{99}

\bibitem{horodeccy} R. Horodecki, P. Horodecki, M. Horodecki, K. Horodecki, Rev. Mod. Phys. {\bf81}, 865 (2009).
\bibitem{amico} L. Amico, R. Fazio, A. Osterloh, V. Vedral, Rev. Mod. Phys. {\bf80}, 517 (2008).
\bibitem{barenco} A. Barenco, C.H. Bennett, R. Cleve, D.P. DiVincenzo, N. Margolus, P. Shor, T.
    Sleator, J.A.  Smolin, and H. Weinfurter, Phys. Rev. A
    {\bf52}, 3457 (1995).
\bibitem{burkard} G. Burkard, D. Loss, D. P. DiVincenzo, and J. A. Smolin, Phys. Rev. B {\bf60}, 11404 (1999).
\bibitem{vinzenzo} D. P. DiVincenzo, D. Bacon, J. Kempe, G. Burkard, and K. B. Whaley, Nature (London) {\bf408}, 339 (2000).
\bibitem{Weinstein} Y. S. Weinstein, and C. S. Hellberg, Phys. Rev. A {\bf72}, 022319
    (2005).
\bibitem{yang} C.P. Yang and J. Gea-Banacloche, Phys. Rev. A {\bf63}, 022311 (2001).
\bibitem{dur} W. D\"{u}r, G. Vidal and J. I. Cirac, Phys. Rev. A {\bf62}, 062314 (2000).
\bibitem{dicarlo} Ch. F. Roos, M. Riebe, H. H\"{a}ffner, W. H\"{a}nsel, J. Benhelm, G. P. T. Lancaster, Ch. Becher, F. Schmidt-Kaler, and R. Blatt, Science {\bf304}, 5676 (2004); L. DiCarlo, M. D. Reed, L. Sun, B. R. Johnson, J. M. Chow, J. M. Gambetta, L.
    Frunzio, S. M.  Girvin, M. H. Devoret, and R. J. Schoelkopf, Nature {\bf467}, 574 (2010).
\bibitem{neeley} M. Neeley, R. C. Bialczak, M. Lenander, E. Lucero, M. Mariantoni, A. D. O'Connell, D. Sank, H. Wang, M. Weides, J. Wenner, Y. Yin, T. Yamamoto, A. N. Cleland, and J. Martinis, Nature {\bf467}, 570 (2010).


\bibitem{laird10} E. A. Laird, J. M. Taylor, D. P. DiVincenzo, C. M. Marcus, M. P.
    Hanson, and A. C. Gossard, Phys. Rev. B 82, 075403 (2010).
\bibitem{takakura} T. Takakura, M. Pioro-Ladriere, T. Obata, Y.-S. Shin, R. Brunner, K.
    Yoshida, T. Taniyama,  and S. Tarucha, Appl. Phys. Lett. {\bf97}, 212104 (2010).
\bibitem{gaudreu} L. Gaudreau, G. Granger, A. Kam, G. C. Aers, S. A. Studenikin, P. Zawadzki,
    M. Pioro-Ladriere, Z. R. Wasilewski  and A. S. Sachrajda,  Nature Phys. {\bf8}, 54 (2012).

\bibitem{kato} Y. Kato, R. C. Myers, D. C. Driscoll, A. C. Gossard, J. Levy, and D. D. Awschalom, Science {\bf299}, 1201 (2003).
\bibitem{golovach} V. N. Golovach, M. Borhani, and D. Loss, Phys. Rev. B {\bf74}, 165319 (2006).
\bibitem{laird} E. A. Laird, C. Barthel, E. I. Rashba, C. M. Marcus, M. P. Hanson, and A. C. Gossard, Phys. Rev. Lett. 99, 246601 (2007).
\bibitem{pioro} M. Pioro-Ladriere, T. Obata, Y. Tokura, Y. S. Shin, T. Kubo, K. Yoshida, T. Taniyama, S. Tarucha, Nature Physics {\bf4}, 776 (2008).

\bibitem{gupta} J. A. Gupta, R. Knobel, N. Samarth, and D. D. Awschalom, Science {\bf292}, 2458 (2001).
\bibitem{imamoglu} A. Imamoglu, D. D. Awschalom, G. Burkard, D. P. DiVincenzo, D. Loss, M.
    Sherwin,  and A. Small, Phys. Rev. Lett. {\bf83}, 4204 (1999).
\bibitem{hanson} R. Hanson, L. P. Kouwenhoven, J. R. Petta, S. Tarucha, and L. M. K.
    Vandersypen, Rev.  Mod. Phys. {\bf79}, 1217 (2007), and references therein.
\bibitem{choi} K.-Y. Choi, Z. Wang, H. Nojiri, J. van Tol, P. Kumar, P. Lemmens, B. S. Bassil, U. Kortz, and N. S. Dalal, Phys. Rev. Lett. {\bf108}, 067206 (2012).


\bibitem{pauncz} R. Pauncz, {\it The Construction of Spin Eigenfunctions}, (Kluwer
    Academic/Plenum Publisher, New York, 2000).

\bibitem{loss} B. R\"{o}thlisberger, J. Lehmann, D.S. Saraga, P. Traber and D. Loss, Phys Rev. Lett. {\bf100}, 100502 (2008).

\bibitem{wang} X. Wang, Phys. Rev. A {\bf64}, 012313 (2001).
\bibitem{miyake} A. Miyake, Phys. Rev. A {\bf67}, 012108 (2003).

\bibitem{Acin} A. Acin, D. Bru{\ss}, M. Lewenstein, and A. Sanpera, Phys. Rev. Lett. {\bf87},
    040401 (2001).
\bibitem{Sabin} C. Sabin and G. Garcia-Alcaine, Eur. Phys. J. D {\bf48}, 435 (2008).

\bibitem{coffman} V. Coffman, J. Kundu and W. K Wootters, Phys. Rev. A {\bf61}, 052306 (2000).
\bibitem{bennett} C. H. Bennett,, D. P. DiVincenzo, J. A. Smolin, and W. K. Wootters,
    Phys. Rev. A {\bf54}, 3824 (1996).

\bibitem{bulka1} B.R. Bu{\l}ka, T. Kostyrko, J. {\L}uczak, Phys. Rev. B {\bf83}, 035301 (2011).
\bibitem{kostyrko} T. Kostyrko and B.R. Bu{\l}ka, Phys. Rev. B {\bf84}, 035123
    (2011).
\bibitem{dassarma} V. W. Scarola, and S. Das Sarma, Phys. Rev. A {\bf71}, 032340 (2005).

\bibitem{Wen} X. G. Wen, F. Wilczek, and A. Zee, Phys. Rev. B {\bf39}, 11413 (1989).
\bibitem{Katsura} H. Katsura, N. Nagaosa, and A. V. Balatsky, Phys. Rev. Lett. {\bf95}, 057205 (2005).
\bibitem{Motrunich} O. I. Motrunich, Phys. Rev. B {\bf73}, 155115 (2006).
\bibitem{Bulaevskii} L. N. Bulaevskii, C. D. Batista, M.V.Mostovoy, and D. I.Khomskii, Phys.
    Rev. B {\bf78}, 024402 (2008).
\bibitem{trif} M. Trif, F. Troiani, D. Stepanenko and D. Loss, Phys. Rev. Lett. {\bf101},
    217201 (2008); M. Trif, F. Troiani, D. Stepanenko and D. Loss, Phys. Rev. B {\bf82}, 045429 (2010).


\bibitem{Hsieh} C.-Y. Hsieh, and P. Hawrylak, Phys. Rev. B {\bf82}, 205311 (2010).

\bibitem{2qd} see for example: J. R. Petta, A. C. Johnson, J. M. Taylor, E. A. Laird, A. Yacoby,
M. D. Lukin, C. M. Marcus, M. P. Hanson, and A. C. Gossard, Science {\bf309}, 2180 (2005);
F. H. L. Koppens, C. Buizert, K. J. Tielrooij, I. T. Vink, K. C. Nowack, T. Meunier, L. P. Kouwenhoven, and L. M. K. Vandersypen, Nature (London) {\bf442}, 766 (2006);
R. Hanson, L. P. Kouwenhoven, J. R. Petta, S. Tarucha, and L. M. K. Vandersypen, Rev. Mod. Phys. {\bf79}, 1217 (2007);
M. Pioro-Ladriere, T. Obata, Y. Tokura, Y. S. Shin, T. Kubo, K. Yoshida, T. Taniyama, and S. Tarucha, Nature Phys. {\bf4}, 776 (2008).


\end{thebibliography}
\end{document}